\pdfoutput=1

\documentclass[11pt,twoside,a4paper,cmspaper,final,collab]{cms-tdr}
\def\svnVersion{486915P}\def\cmsCernNoTag{CERN-EP-2017-268}\def\cmsCernDate{\today}\def\cmsMessage{Published in Physics Letters B as \href{http://dx.doi.org/10.1016/j.physletb.2018.11.063}{\doi{10.1016/j.physletb.2018.11.063.}}}

\begin{document}\cmsNoteHeader{HIN-16-019}

\hyphenation{had-ron-i-za-tion}
\hyphenation{cal-or-i-me-ter}
\hyphenation{de-vices}
\RCS$HeadURL: svn+ssh://svn.cern.ch/reps/tdr2/papers/HIN-16-019/trunk/HIN-16-019.tex $
\RCS$Id: HIN-16-019.tex 486880 2019-01-22 14:50:53Z jcastle $
\newlength\cmsFigWidth
\ifthenelse{\boolean{cms@external}}{\setlength\cmsFigWidth{0.98\columnwidth}}{\setlength\cmsFigWidth{0.49\textwidth}}
\ifthenelse{\boolean{cms@external}}{\providecommand{\cmsLeft}{top\xspace}}{\providecommand{\cmsLeft}{left\xspace}}
\ifthenelse{\boolean{cms@external}}{\providecommand{\cmsRight}{bottom\xspace}}{\providecommand{\cmsRight}{right\xspace}}
\providecommand{\rootsNN}{\ensuremath{\sqrt{\smash[b]{s_{_\text{NN}}}}}\xspace}
\cmsNoteHeader{HIN-16-019}
\title{Non-Gaussian elliptic-flow fluctuations in PbPb collisions at $\sqrt{s_{_\text{NN}}} = 5.02$\TeV}

\date{\today}

\abstract{
Event-by-event fluctuations in the elliptic-flow coefficient
$v_2$ are studied in PbPb collisions at $\rootsNN = 5.02$\TeV
using the CMS detector at the CERN LHC. Elliptic-flow probability
distributions ${p}(v_2)$ for charged particles
with transverse momentum $0.3 < \pt < 3.0$\GeVc and pseudorapidity $\abs{ \eta } < 1.0$
are determined for different collision centrality classes.
The moments of the ${p}(v_2)$ distributions are used
to calculate the $v_{2}$ coefficients based on cumulant orders 2, 4, 6, and 8.
A rank ordering of the higher-order cumulant results and nonzero standardized skewness values
obtained for the ${p}(v_2)$ distributions indicate non-Gaussian initial-state fluctuations.
Bessel--Gaussian and elliptic power fits to the
flow distributions are studied to characterize the initial-state spatial anisotropy.
}

\hypersetup{%
pdfauthor={CMS Collaboration},%
pdftitle={Non-Gaussian elliptic-flow fluctuations in PbPb collisions at sqrt(sNN) = 5.02 TeV},%
pdfsubject={CMS},%
pdfkeywords={Event-by-Event Elliptic Flow, Non-Gaussian Flow Fluctuations, Unfolding}}

\maketitle

\section{Introduction}
\label{sec:intro}
Ultrarelativistic heavy ion collisions at both the BNL Relativistic Heavy Ion Collider (RHIC) and the CERN Large Hadron Collider (LHC) create a hot and dense state of matter that consists of strongly interacting quarks and gluons, the ``quark-gluon plasma"  (QGP)~\cite{Arsene:2004fa, Adcox:2004mh, Back:2004je, Adams:2005dq, Aad:2010bu, Chatrchyan:2011sx,Aamodt:2010jd}.  Measurements of azimuthal particle correlations resulting from these collisions reveal properties of the QGP, but also of the initial state of a heavy-ion collision.  In particular, the overall shape and fluctuations in the initial-state transverse energy density  transformed by the hydrodynamic evolution of the medium into anisotropies in the final-state momentum space for the emitted particles~\cite{Ollitrault:1992bk, Alver:2010gr,Gronqvist:2016hym}, as reflected in the azimuthal charged-particle density. The early RHIC measurements of the  azimuthal correlations showed that the QGP could be described well by hydrodynamic models~\cite{Heinz:2011kt}, with a shear viscosity to entropy density ratio ($\eta/s$) that is of the order of the lowest possible value for a quantum fluid~\cite{Policastro:2001yc, Kovtun:2004de}.

The azimuthal charged-particle density can be characterized by a Fourier expansion, with
\begin{linenomath}
  \begin{equation}
    \frac{\rd{}N_\text{ch}}{\rd\phi} \propto 1 + 2 \sum_{n=1}^{\infty}  v_{n} \cos{ [n \left(\phi -  \Phi_{n} \right)] }.
    \label{eqn:dndphi}
  \end{equation}
\end{linenomath}
Here, the $n$th-order flow vector for a given event is ${\vec v_n} \equiv \left( {{v_{n} \cos{ \Phi_{n}}}, {v_{n}\sin{ \Phi_{n}}} } \right)$, where  $\Phi_{n}$ is the angle of the intrinsic $n$th-order flow symmetry plane, as determined by the geometry of the participant nucleons. The experimentally accessible ``event plane'' angle, $\Psi _n^{{\rm{obs}}}$, is based on the direction of maximum outgoing particle density and is, on average,  in the same direction as $\Phi_n$, but fluctuates about $\Phi_n$ because of resolution effects due to finite particle multiplicities.

By calculating the flow coefficients over a large number of events,  the underlying probability distribution functions of individual Fourier coefficients can be determined.  While the mean values of the $v_{n}$ distributions can be related to the overall shape of the interaction region, the higher order moments can be used to  constrain the origin and the nature of the initial-state fluctuations and help disentangle the initial-state effects from the subsequent evolution of the medium~\cite{Yan:2014afa,Yan:2014nsa}.  Here, an event-by-event analysis is performed where it is possible to reduce the sensitivity of the results to nonflow correlations~\cite{Aad:2013xma} and to clearly establish higher-order moments of the  $n=2$ (elliptic) distribution function.   The mean of this distribution, $\left\langle {{v_2}} \right\rangle $, is largely determined by the lenticular shape of the collision overlap region.

While the final-state particle distribution is characterized by the ${\vec v_n}$ coefficients,
the initial-state spatial anisotropy can be characterized by a harmonic expansion in terms of
 eccentricity vectors $\vec{\varepsilon}_n$~\cite{Voloshin:2008dg,Teaney:2010vd,Qiu:2011iv,Gardim:2011xv}.
For a given impact parameter, fluctuations in the initial-state
transverse energy density lead to event-by-event differences in the orientation and
magnitude of the $\vec{\varepsilon}_n$ vectors with respect to the experimentally inaccessible ``reaction plane,'' defined by the collision
impact parameter and beam directions. The presence of a
nonzero viscosity will degrade the correspondence between initial- and final-state
anisotropies~\cite{Song:2012ua, Heinz:2011kt}. Still, an almost linear dependence
is expected for the lowest order $n = 2$~\cite{Kolb:2000sd,Bhalerao:2005mm,Bhalerao:2006tp,Alver:2008zza,Gombeaud:2009ye}  and $n = 3$~\cite{Alver:2010gr,Teaney:2010vd} harmonics, with
$v_n = k_n~\varepsilon_n$~\cite{Qiu:2011iv}.  Here, $v_n \equiv \abs{\vec v_n}$,
 $\varepsilon_n \equiv \abs{\vec{\varepsilon_n}}$, and $k_n$ is the flow response coefficient.
The probability distribution functions of the magnitudes of the
$\vec{\varepsilon}_n$ vectors,  $p(\varepsilon_n)$,
can be related to the corresponding $p(v_n)$ distribution assuming a linear response, according to:
\begin{linenomath}
  \begin{equation}
    \label{eq:3_vlk}
    p(v_n) = \frac{\rd\varepsilon_n}{\rd{}v_n}~p(\varepsilon_n) = \frac{1}{k_n}~p\Biggl(\frac{v_n}{k_n}\Biggr),
  \end{equation}
\end{linenomath}
where the  $k_n$ term is expected to depend on the hydrodynamic evolution of the medium~\cite{Voloshin:1999gs,Drescher:2007cd}.

The  elliptic-flow $p(v_2)$ distribution can be characterized using
the experimentally determined multiparticle cumulant flow harmonics $v_2\{m\}$~\cite{Borghini:2000sa, Borghini:2001vi},
where $m$ is the cumulant
order.  Alternatively,
the  distribution can be determined directly, as shown by the
ATLAS Collaboration~\cite{Aad:2013xma} and as done here, by removing finite-multiplicity resolution effects
in the measured $p(v_2^\text{obs})$ distribution through an unfolding technique.
The cumulant harmonics are expressed in terms of the moments of the $p(v_2)$ distribution~\cite{Voloshin:2007pc,Jia:2014jca}:
\begin{linenomath*}
\begin{equation}
  \begin{aligned}
      v_{2}\{ 2 \}^2 \equiv&  \text{E}( v_2^2 ), \\
      v_{2}\{ 4 \}^4 \equiv&  -\text{E}( v_2^4 ) + 2 \text{E}(v_2^2 )^2, \\
      v_{2}\{ 6 \}^6 \equiv&  \left( \text{E}( v_2^6 ) - 9
      \text{E}( v_n^4 ) \text{E}( v_2^2 ) + 12 \text{E}( v_2^2 )^3 \right) / 4, \\
      v_{2}\{ 8 \}^8 \equiv&  {-}( \text{E}( v_2^8 ) - 16 \text{E}( v_2^6 ) \text{E}( v_2^2 ) - 18 \text{E}( v_2^4 )^2 \\
     & + 144 \text{E}( v_2^4 ) \text{E}( v_2^2 )^2 - 144 \text{E}( v_2^2 )^4
     ) /33,
\label{eqn:cumuOrders}
\end{aligned}
\end{equation}
\end{linenomath*}
where $\text{E}(v_2^k) \equiv \int v_2^k p(v_2) \rd{}v_2$.
The unitless standardized skewness of a probability distribution is a measure of
the  asymmetry about its mean. For the case of elliptic flow, the standardized
skewness with respect to the reaction plane can be estimated using the cumulant flow harmonics as in Ref.~\cite{Giacalone:2016eyu}:
\begin{linenomath}
  \begin{equation}
    \gamma_1^{\text{exp}} \equiv -6\sqrt{2}v_2\{4\}^2 \frac{v_2\{4\} - v_2\{6\}}{\Bigl( v_2\{2\}^2 - v_2\{4\}^2\Bigr)^{3/2}}.
    \label{eqn:gamma1exp}
  \end{equation}
\end{linenomath}
Hydrodynamic calculations find this estimate to be in good agreement with
the actual skewness except for the most peripheral events~\cite{Giacalone:2016eyu}.

The standardized skewness estimate vanishes for fluctuations that arise from an isotropic Gaussian transverse initial-state energy density profile. In this case, the  ${p}(v_{2})$ distribution is found by taking an integral over the azimuthal dependence of the two-dimensional Gaussian function~\cite{Voloshin:2007pc, Voloshin:1994mz}. The resultant, one-dimensional distribution has a Bessel-Gaussian shape, where the even cumulant coefficients $v_2\{ m \}$  with $m \geq 4$ are degenerate~\cite{Voloshin:2007pc}. The observation for PbPb collisions that
$v_{2}\{4\} \approx v_{2}\{6\} \approx v_{2}\{8\}$~\cite{Khachatryan:2015waa, Abelev:2014mda, Aad:2014vba}, where the approximate equalities are within a few percent, suggests that the ${\vec v_2}$ fluctuations can be well described by a two-dimensional Gaussian function~\cite{Voloshin:2007pc}.

Still, non-Gaussian fluctuations are expected in the initial-state energy density~\cite{Giacalone:2016eyu}, which should lead to differences in the higher order cumulant coefficients.  Such differences have been reported by the ATLAS Collaboration~\cite{Aad:2013xma} in a similar measurement  of peripheral PbPb collisions to that reported here. The precision of the LHC measurements allows for these differences to be explored in detail, giving a new method to investigate the initial-state behavior. The elliptic power function has been suggested to describe the asymmetric behavior of the ${p}(\varepsilon_n)$ distributions~\cite{Yan:2014op,Yan:2014afa,Yan:2014nsa}, noting that the Bessel-Gaussian distribution reproduces neither Glauber Monte Carlo nor IP-Glasma results other than for very central events~\cite{Yan:2014afa}. This function is based on the assumption that the initial energy density profile of the collision is a superposition of $N$ point-like, independent sources. In terms of the harmonic-flow coefficients and assuming a linear response,
\begin{linenomath}
  \begin{equation}
  \label{eq:1_vlk}
    p(v_n) = \frac{2 \alpha v_n}{\pi k_n^2}
    (1-\varepsilon_0^2)^{\alpha+1/2}
    \int_0^{\pi} \frac{(1-v_n^2/k_n^2)^{\alpha-1}\rd\phi}
    {(1-\varepsilon_0 v_n\cos{\phi}/k_n)^{2\alpha+1}},
  \end{equation}
\end{linenomath}
where $\varepsilon_{0}$ is approximately equal to the mean eccentricity in the reaction plane and $\alpha$, which is approximately proportional to $N$, describes the size of the eccentricity fluctuations. The elliptic power distribution reduces to a Gaussian, Bessel-Gaussian,  or power distribution form with the appropriate choice of parameters~\cite{Yan:2013laa} and has the advantage of naturally incorporating the  unit constraint on  eccentricity, where $\abs{\epsilon_{n}} < 1$.

In this Letter,  the  $p(v_2)$ distributions for charged particles
in the  pseudorapidity range  $\abs{\eta} < 1.0$ and with transverse
momenta $0.3 < \pt < 3.0$\GeVc are presented for PbPb collisions at $\rootsNN = 5.02$\TeV
collected with the CMS detector at the LHC. The results are shown in bins of centrality,
defined as fractions of the total inelastic hadronic cross section, where
0\% corresponds to the events with the greatest hadronic activity in the forward direction ($\abs{\eta} > 3.0$).
The elliptic-flow harmonic values for different cumulant orders are determined based on the moments of the
$p(v_2)$ distributions, with these results used to estimate the standardized skewness of the flow distribution.
Elliptic power and Bessel--Gaussian fits to the  flow distributions
are presented to gain further insight into the initial-state and its fluctuations.

\section{The CMS detector}
\label{sec:CMSDetector}
The central feature of the CMS apparatus is a superconducting solenoid of
6\unit{m} internal diameter, providing a magnetic field of 3.8\unit{T}. Within
the solenoid volume are a silicon pixel and strip tracker, a lead tungstate
crystal electromagnetic calorimeter, and a brass and scintillator hadron
calorimeter, each composed of a barrel and two endcap sections. Muons are detected in gas-ionization chambers embedded in
the steel flux-return yoke outside the solenoid.

The barrel and endcap detectors provide coverage in the range $\abs{\eta}<3.0$, with
Hadron Forward calorimeters (HF) extending the pseudorapidity coverage to $3.0 < \abs{\eta} < 5.2$.
The HF detectors  are
used both to select events for the analysis and to determine the collision centrality. The HF calorimeters
are azimuthally subdivided into $20^{\circ}$ modular wedges and further segmented to
form $0.175\times10^{\circ}$ $(\Delta\eta{\times}\Delta\phi)$ towers.
The silicon tracker measures charged particles within the
range $\abs{\eta} < 2.5$. It consists of 1440 silicon pixel and
15\,148 silicon strip detector modules. At midrapidity, there are 3 pixel detector layers and 10 strip detector layers.  At the outer edge of the tracker acceptance, there are 2 pixel detector layers and 12 strip detector layers.  For nonisolated particles
of $1 < \pt < 10$\GeVc and $\abs{\eta} < 1.4$, the track resolutions are
typically 1.5\% in \pt and 25--90 (45--150)\unit{\mum} in the transverse
(longitudinal) distance of closest approach~\cite{TRK-11-001}. A more detailed
description of the CMS detector, together with a definition
of the coordinate system used and the relevant kinematic variables,
can be found in Ref.~\cite{Chatrchyan:2008zzk}.

\section{Event and track selection}
\label{sec:evtAndTrkSel}
This analysis is based on a PbPb minimum bias data set with $\rootsNN = 5.02$\TeV and corresponding to an integrated luminosity of 26\mubinv,
collected in 2015 .
The minimum-bias trigger used  requires coincident
signals in the HF calorimeters at both ends of the CMS detector with
energy deposits above a predefined  energy threshold of approximately 1\GeV and
the presence of both colliding bunches at the interaction point as determined using
beam pickup timing monitors. By requiring  colliding
bunches, events due to noise
(\eg, cosmic rays and beam backgrounds) are largely suppressed.
Events are further selected offline by requiring at least three
towers with an energy above 3\GeV
in each of the two HF calorimeters. The primary vertex for each event is
chosen as the reconstructed vertex with the largest number of associated
tracks. Primary vertices are required to have at least
two associated tracks and to be located within 15\,(0.2)\unit{cm} of the nominal
collision point along the longitudinal (transverse) direction. To suppress
contamination from events with multiple collisions in the same bunch crossing (pileup), the procedure outlined in
Ref.~\cite{Chatrchyan:2011sx} is followed. Here, compatibility scores
based on the number of pixel clusters with widths compatible with particles
originating from each primary vertex are determined and events with primary
vertices with compatibility scores below a predefined threshold are rejected
as pileup. After applying the selection criteria, the average number of collisions per bunch crossing is
less than $\approx$0.001 for the events used in this analysis, with a pileup fraction $<$0.05\%.

Track reconstruction~\cite{TRK-11-001,Khachatryan:2016odn} is performed in
two iterations to ease the computational load for high-multiplicity central
PbPb collisions. The first iteration reconstructs tracks from signals
(``hits'') in the silicon pixel and strip detectors compatible with a
trajectory of $\pt>0.9$\GeVc. These tracks are required to have consistency with originating from the primary vertex, having a longitudinal association
significance ($d_{z}/\sigma_{d_{z}}$) and a distance of closest approach
significance ($d_{0}/\sigma_{d_{0}}$) each less than 3. In addition, the
\pt resolution~\cite{TRK-11-001,Khachatryan:2016odn} for each track, $\sigma_{\pt}/\pt$, is
required to be less than 10\% and tracks are required to have at least 11
out of the possible 14 hits along their trajectory in the pixel and strip trackers.  To reduce the
number of misidentified tracks, which can occur when the hit pattern is consistent with more than one possible track solution, the chi-squared per degree of freedom, $\chi^2/\mathrm{dof}$, associated with fitting the
track trajectory through the different pixel and strip layers
must be less than 0.15 times the total number of layers with hits along
the trajectory of the track. The second iteration reconstructs tracks compatible
with a trajectory of $\pt>0.2$\GeVc using solely the pixel detector. These tracks
are required to have longitudinal association significance
$d_{z}/\sigma_{d_{z}} < 8$ and a fit $\chi^2/\mathrm{dof}$ value less than
12 times the number of layers with hits along the trajectory of the track.  In the final analysis,
first iteration tracks with $\pt > 1.0$\GeVc are used together with pixel-detector-only tracks with $\pt < 2.4$\GeVc
after removing duplicates. Track reconstruction for the merged iterations
has a combined geometric acceptance and efficiency exceeding 60\% for
$\pt \approx 1.0$\GeVc and $|\eta| < 1.0$. When the track \pt is below 1\GeVc,
the acceptance and efficiency steadily drops, reaching approximately 40\%
at $\pt \approx 0.3$\GeVc.

\section{Analysis technique}
\label{sec:analysis}

Analyses of  flow harmonics using  multiparticle cumulants  were initially introduced as a way to minimize nonflow effects~\cite{Borghini:2001vi}.
These analyses have been based on either the generating function formalism~\cite{Borghini:2001vi} or, more recently, through direct calculation~\cite{Bilandzic:2013kga}. The unfolding procedure employed here, as introduced by the ATLAS collaboration~\cite{Aad:2013xma}, is expected to give similar results to a multiparticle cumulant analysis, but with reduced sensitivity to multiplicity fluctuations and nonflow effects~\cite{Jia:2013tja}.

The event-by-event $v_2$ coefficients and phases in Eq.~(\ref{eqn:dndphi}) can
be estimated with
\begin{linenomath}
  \begin{equation}
    \begin{split}
      v_{2,x}^{\text{obs}} &= \abs{\vec{v}_{2}^{\text{\,obs}}}\cos{(2 \Psi_{2}^{\text{obs}})} = \langle \cos{(2 \phi)} \rangle = \frac{\sum_i w_i \cos{(2 \phi_i)}}{\sum_i w_i}, \\
      v_{2,y}^{\text{obs}} &= \abs{\vec{v}_{2}^{\text{\,obs}}}\sin{(2 \Psi_{2}^{\text{obs}})} = \langle \sin{(2 \phi)} \rangle = \frac{\sum_i w_i \sin{(2 \phi_i)}}{\sum_i w_i}, \\
      \abs{\vec{v}_{2}^{\text{obs}}} &= \sqrtsign{ \left ( v_{2,x}^{\text{obs}} \right )^2 + \left ( v_{2,y}^{\text{obs}} \right )^2},
    \end{split}
    \label{eqn:ebyeVn}
  \end{equation}
\end{linenomath}
where $\phi_i$ is the azimuthal angle of the track,  $\Psi_{2}^{\text{obs}}$ is the event plane angle for the 2nd harmonic,
the angular brackets denote an efficiency weighted average over all particles in a
given range of phase space for an event, and $w_i = 1 / \varepsilon_i$ is the inverse
of the tracking efficiency $\varepsilon_i\left(\pt, \eta \right)$ of the $i^\text{th}$ track.
The analysis does not require the explicit calculation of the event plane angle for each event.
In the absence of particle correlations unrelated to the hydrodynamic flow behavior (``nonflow''),
the observed event-by-event flow vectors of Eq.~(\ref{eqn:ebyeVn})
will approach the true  underlying flow vectors as the particle multiplicity becomes large.
In addition to the efficiency weighting, a standard recentering procedure~\cite{Poskanzer:1998yz}, where the event average x- and y-components
of the flow vector are required to equal zero,
is applied to further suppress acceptance biases.

Events are sorted into different centrality classes, as determined by the transverse
energy deposited in the HF calorimeters~\cite{Chatrchyan:2011sx}, and the magnitudes of the estimated flow
vectors are used to construct the ``observed'' ${p}(v_{2}^{\text{obs}})$ distributions
for each class. Finite particle multiplicities result in a statistical fluctuation
of the $v_2^{\text{obs}}$ estimate for a given event about the true underlying $v_2$
value by a response function ${p}(v_2^{\text{obs}}| v_2)$.
This, in turn, results in a ${p}(v_2^{\text{obs}})$ distribution
that is broader than the underlying ${p}(v_2)$
behavior. The observed distribution can  be expressed
as a convolution of the underlying flow behavior and the
response function
\begin{linenomath}
  \begin{equation}
    \label{eqn:FlowUnfolding}
    {p}(v_{2}^{\text{obs}}) = {p}(v_{2}^{\text{obs}} | v_{2}) \ast {p}(v_{2}).
  \end{equation}
\end{linenomath}

\noindent A data-based technique, first introduced by the ATLAS Collaboration~\cite{Aad:2013xma},
was used to build the response function in Eq.~(\ref{eqn:FlowUnfolding}).
This technique divides the full event sample into two symmetric subevents ($a$ and $b$) based on pseudorapidity.
Given that $v_2(\eta)$ is symmetric about
$\eta=0$ on average for the symmetric PbPb system, the physical flow signal cancels in the distribution
of flow vector differences from each subevent $p(\vec{v}_n^a - \vec{v}_n^b)$. The resulting distribution
contains residual effects from multiplicity-related fluctuations and nonflow
effects~\cite{Jia:2013tja} and provides a basis for building the response function. The ability of the analysis procedure to suppress nonflow effects was
studied by introducing a $v_2$ signal on top of \HIJING\ 1.383~\cite{Gyulassy:1994ew} simulated events, which contain nonflow.
The EbyE analysis is found to recover the ``truth'' to within 0.1\%.

\begin{figure*}[!htbp]
  \centering
    \includegraphics[width=1.0\textwidth]{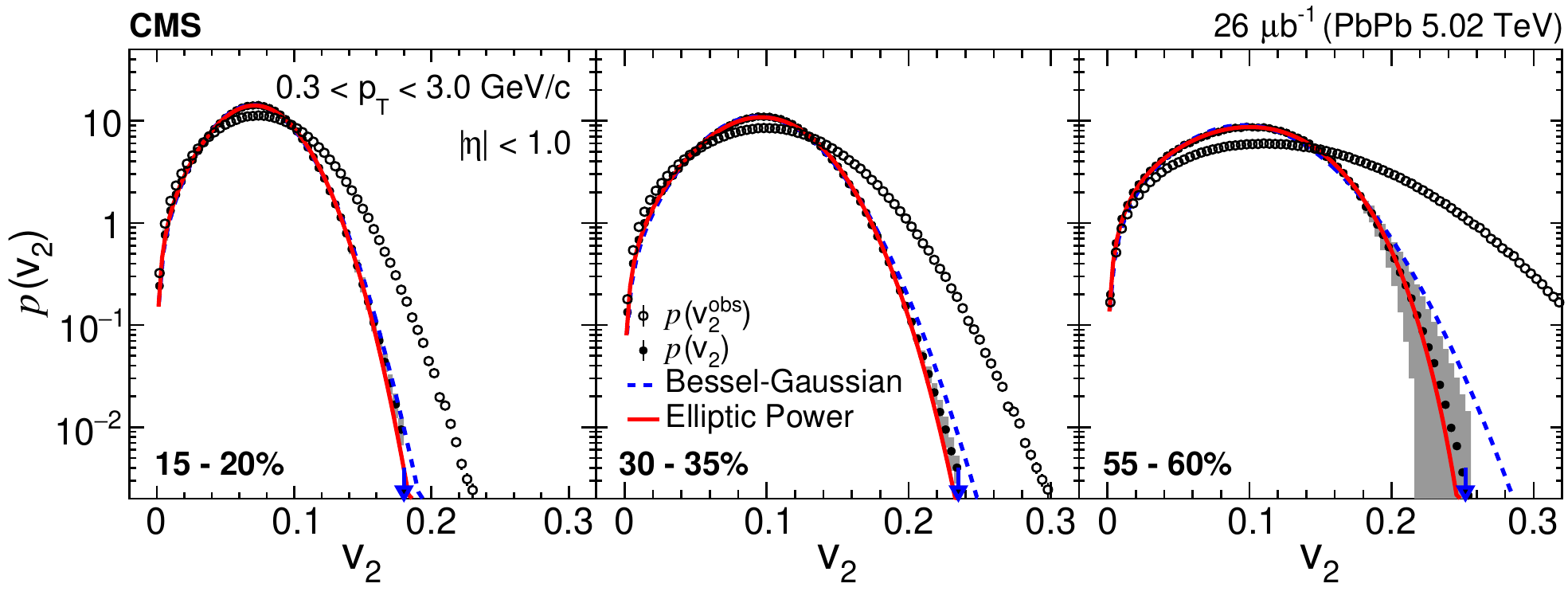}
    \caption{Representative final unfolded $p(v_2)$ distributions (closed black circles)
             in three centrality bins (15--20\%, 30--35\%, and 55--60\%) obtained
             using D'Agostini iteration unfolding. Respective observed $p(v_2^\text{obs})$ distributions
             (open black squares) are shown to illustrate the statistical resolution present in each centrality bin
             prior to unfolding. Systematic uncertainties from the unfolding
             procedure are presented as shaded bands. Distributions are fitted with Bessel--Gaussian
             (dashed blue lines) and elliptic power (solid red lines) functions to infer information
             on the underlying $p(\varepsilon_2)$ distributions. The vertical blue arrows indicate the $\langle v_2 \rangle + 4\sigma_{v_2}$ cutoff discussed in the text.}
    \label{fig:resultsUnf}
\end{figure*}

To unfold the  effects of multiplicity-related fluctuations, the D'Agostini iterative method with early stopping (regularization)
~\cite{D'Agostini1995487, richardson1972bayesian, lucy1974iterative} was used to obtain
a maximum likelihood estimate of the underlying
$p(v_2)$ behavior. The analysis was done using the \textsc{RooUnfold}~\cite{Adye:2011gm} package of the  \textsc{root}
data analysis framework~\cite{Brun:1997pa}.
The unfolding procedure becomes increasingly sensitive to statistical fluctuations when
the number of iterations is allowed to run to large values, resulting
in unphysical oscillations in the low event count tails of the unfolded
distribution. The regularization criterion used
to suppress these oscillations is to apply the response function to
each unfolding iteration (``refolding'') and compare the resulting distribution to the observed
one.  Iterations are stopped when the $\chi^2/\mathrm{dof}$ between the refolded and
observed distribution is approximately equal to one.   After this final unfolding iteration is reached,
the resulting distribution is truncated above
$\langle v_2 \rangle + 4\sigma_{v_2}$ to further suppress any residual artifacts
in the tails that result from the unfolding procedure. Representative final unfolded distributions are shown
in Fig.~\ref{fig:resultsUnf}. In addition, ${p}(v_2^{\text{obs}})$  distributions are plotted for each centrality
to illustrate the statistical resolution effects present prior to unfolding. The fits shown in Fig.~\ref{fig:resultsUnf}
are discussed in Section~\ref{sec:results}.

\section{Systematic uncertainties}
\label{sec:systematics}
A number of potential sources of systematic uncertainties for the $v_2\{m\}$ values extracted from the unfolded $p(v_2)$ distributions were considered.
The systematic uncertainties that arise from the vertex $z$ position were investigated by splitting the default vertex range into two windows of $\abs{z_\text{vtx}} < 3.0$\unit{cm} and $3.0 < \abs{z_\text{vtx}} < 15.0$\unit{cm} and comparing the results from the two ranges. The resulting uncertainties range from 5\% for central events, decreasing to 0.5\% for mid-central events.
To estimate the bias from misidentified tracks, the track quality criteria described in Section~\ref{sec:evtAndTrkSel} were varied. Two scenarios were considered, with one  increasing  and the other decreasing the probability of misidentifying a track. The results of these two scenarios were compared to the values obtained in the default analysis.  The resulting uncertainties range from 2\% for central events to 1\% for mid-central events.
To estimate the systematic uncertainty in the choice of response function, the unfolding procedure was repeated using an analytic response function obtained from a Gaussian fit to the data-driven statistical resolution distribution~\cite{Aad:2013xma}. The resulting uncertainties are 3\% for central events and decrease to 1\% for mid-central events.
Other sources of potential systematic bias were explored and found to be negligible.  To assess the potential bias from residual pileup events, the threshold for determining pileup events was raised to decrease the probability of including events with multiple collisions in the analysis.  The bias from unfolding regularization was studied by modifying the $\chi^2/\mathrm{dof}$ goodness-of-fit regularization criteria and comparing the cases when the refolding $\chi^2/\mathrm{dof}$ cutoff is 2.0 relative to when it is 1.0. To test the potential bias that might result from the $4\sigma$ truncation of the final unfolded distributions, the truncation point was varied between $3.5\sigma$ and $4.5\sigma$.  To assess the uncertainty on the choice of the prior, the unfolding was repeated using priors that were systematically transformed to have 10\% larger and smaller means than the default prior. No significant bias was found with these variations of the prior.
The total systematic uncertainties were obtained by adding the contribution from each source in quadrature. The $v_2$ values calculated for the different cumulant orders have a total systematic uncertainty of the order of 5\% for central collisions, which decreases to 1\% in mid-central collisions.

As all of the systematic uncertainties are expected to be correlated between the different cumulant orders, with the same data used in the calculation of each order, all of the above studies were also performed for the ratios of different orders and for the skewness estimate given by Eq.~\ref{eqn:gamma1exp}.
For the ratios, the total systematic uncertainty is found as 1\% for central collisions, decreasing to 0.1\% for mid-central collisions.
The standardized skewness is very sensitive to small fluctuations in the cumulant flow harmonics, resulting in a systematic uncertainty of 100\% for central collisions that reduces to 20\% for mid-central collisions.

\section{Results}
\label{sec:results}
\begin{figure}[!htbp]
  \centering
    \includegraphics[width=0.49\textwidth]{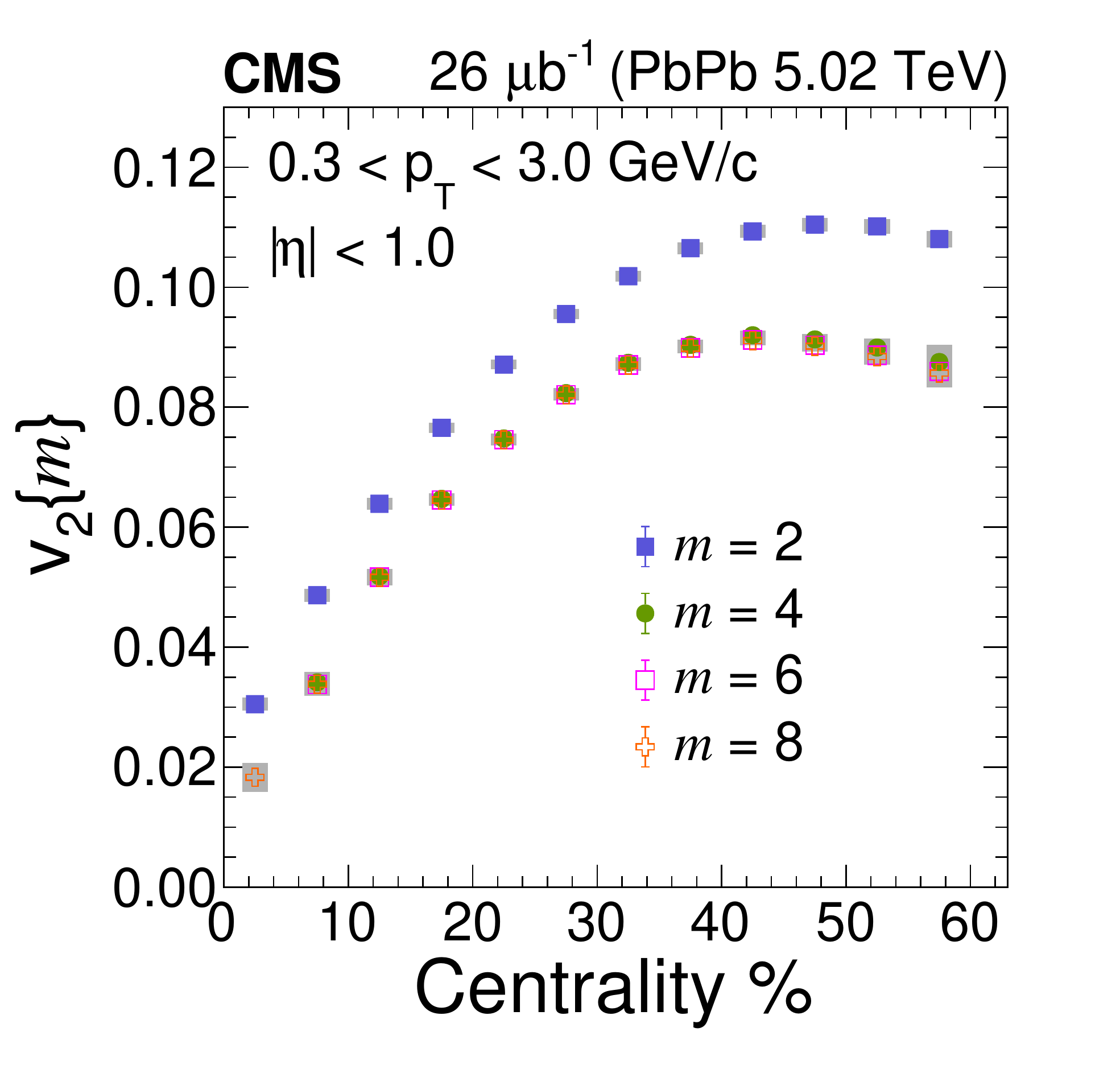}
    \caption{Elliptic-flow cumulant harmonics with values obtained from
             the moments of the unfolded $p(v_2)$ distributions.
             Systematic uncertainties  are shown as gray bands. For most centralities, the uncertainties are smaller than the symbol size.
             }
    \label{fig:resultsCumu}
\end{figure}

\begin{figure*}[!htbp]
   \begin{center}
      \includegraphics[width=1.0\textwidth]{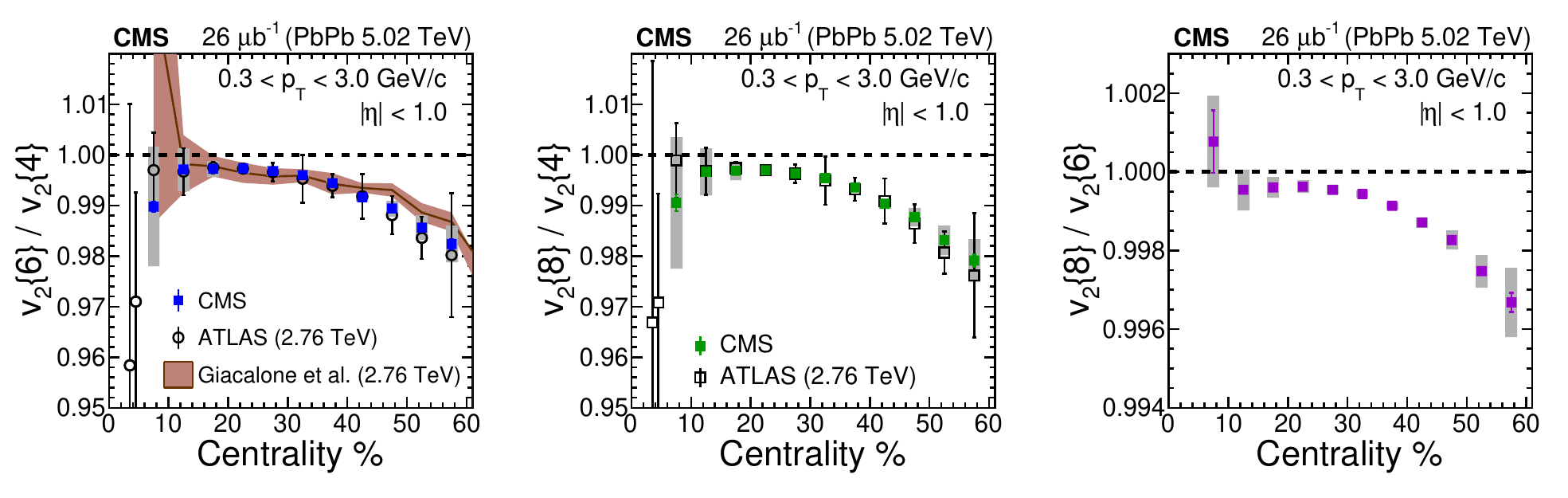}
      \caption{Ratios of higher order cumulant elliptic-flow harmonics with values obtained from the moments
               of the unfolded $p(v_2)$ distributions. Both statistical (lines)
               and systematic (gray bands) uncertainties are shown. Hydrodynamic predictions for
               2.76\TeV collisions from Ref.~\cite{Giacalone:2016eyu} are presented as a dark color
               band and are compared to the measured $v_{2}\{6\}/v_2\{4\}$ ratio. In
               addition, higher order cumulant ratios reported by the ATLAS
               Collaboration for 2.76\TeV collisions~\cite{Aad:2014vba} with $0.5 < \pt < 20.0$\GeVc and $\abs{\eta}<2.5$
               are compared to the 5.02\TeV measurement. The error bars on the ATLAS
               measurement represent the quadratic sum of statistical and systematic
               uncertainties and points are offset horizontally for clarity.}
      \label{fig:cumuRatios}
   \end{center}
\end{figure*}

The cumulant elliptic-flow harmonics obtained from the moments
of the unfolded $p(v_2)$ distributions using Eq.~(\ref{eqn:cumuOrders})
are shown in Fig.~\ref{fig:resultsCumu} for cumulant orders 2, 4, 6, and 8.
It was not possible to obtain 0--5\% central results for $v_2\{4\}$ and $v_2\{6\}$ because the right-hand side of Eq. 3 was found to be negative for these values.   This behavior might be a consequence of volume fluctuations dominating the cumulant behavior for these central events, as discussed in Ref.~\cite{Zhou:2018fxx}.
The cumulant results exhibit the previously observed
$v_2\{2\} > v_2\{4\} \approx v_2\{6\} \approx v_2\{8\}$ behavior.
The centrality-dependent ratios for the elliptic-flow coefficients
obtained for different cumulant orders are shown in
Fig.~\ref{fig:cumuRatios}. For most centrality ranges, the ratios indicate a rank ordering of the cumulants, with differences on the
order of a few percent and with $v_2\{4\} >
v_2\{6\} >  v_2\{8\}$, that is qualitatively inconsistent
with a pure Gaussian fluctuation model of flow harmonics. The differences increase as the collisions become more peripheral.
The calculated $v_{2}\{6\}/v_2\{4\}$ ratio based on an event-by-event hydrodynamic calculation using
Monte Carlo Glauber initial conditions~\cite{Miller:2007ri} and an $\eta/s$ value of 0.08
is shown by the shaded band. This simulation is for pions with
$0.2<\pt<3.0$\GeVc in PbPb collisions at
$\rootsNN= 2.76$\TeV~\cite{Giacalone:2016eyu}.  Also shown are results
from the ATLAS Collaboration~\cite{Aad:2014vba}
for PbPb collisions at  2.76\TeV
and for charged particles with $0.5 < \pt < 20.0$\GeVc and $\abs{\eta}<2.5$.
The calculation is consistent with the experimental results found at both beam energies.
The similarity between experimental results with 2.76 and 5.02\TeV is consistent with the small
changes in the initial-state eccentricities expected between these
energies~\cite{Noronha-Hostler:2015uye} and the expectation that the cumulant flow harmonic ratios follow those of the corresponding eccentricity ratios~\cite{Giacalone:2016eyu}.

\begin{figure}[!htbp]
  \centering
    \includegraphics[width=0.5\textwidth]{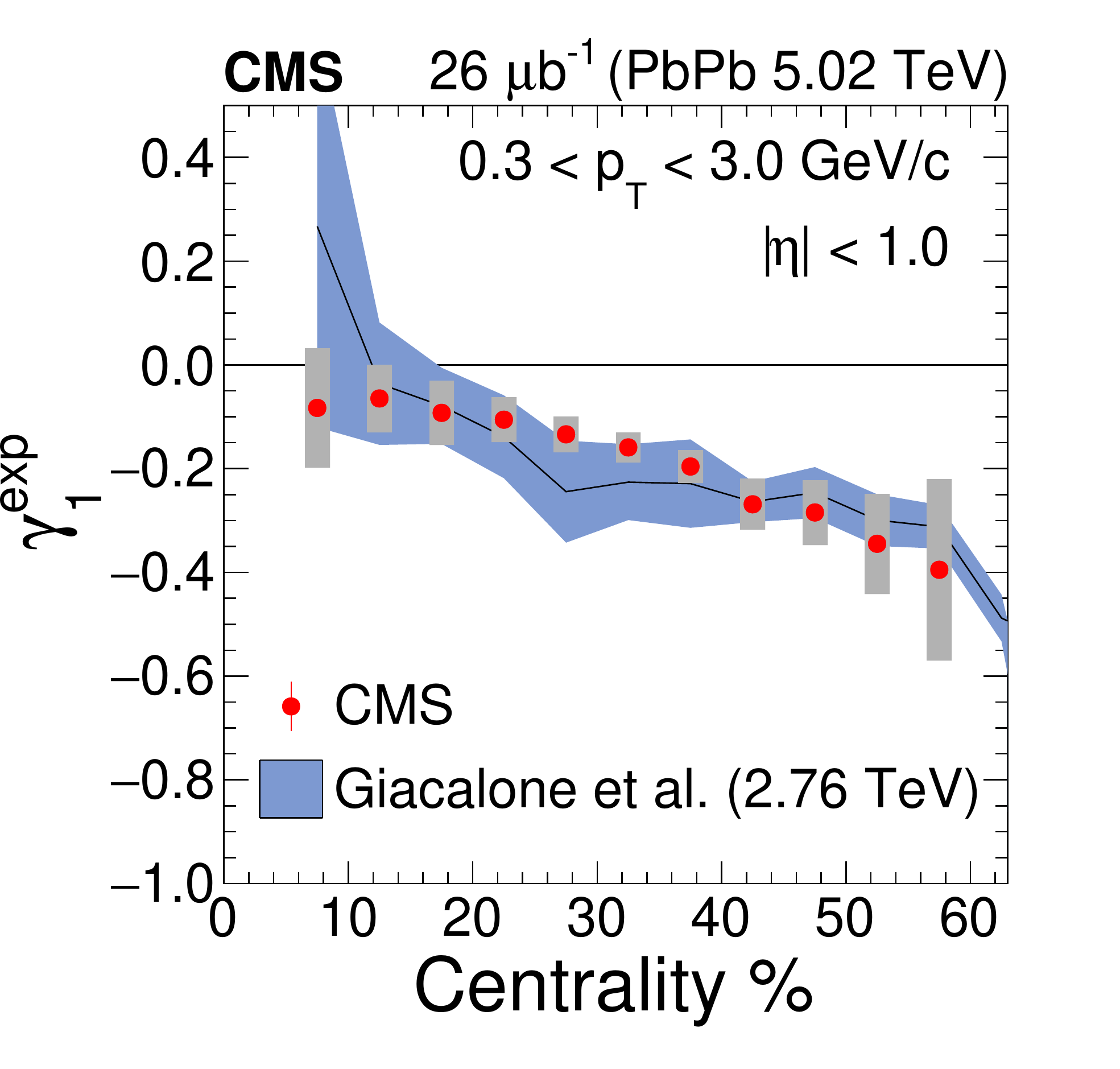}
    \caption{The skewness estimate with respect to the reaction plane determined using the
             elliptic-flow harmonic based on different cumulant orders.
             Both statistical and systematic uncertainties are shown, where statistical uncertainties are
             smaller than the data points. Hydrodynamic model predictions for
             2.76\TeV PbPb collisions from Ref.~\cite{Giacalone:2016eyu} are shown as a
             colored band.}
    \label{fig:resultGamma1Exp}
\end{figure}

Figure~\ref{fig:resultGamma1Exp} shows the centrality dependence of the standardized skewness $\gamma_1^{\text{exp}}$.
Finite values are found for the standardized skewness for collisions with centralities greater than $\approx$15\%.
The hydrodynamic predictions for the $\gamma_1^{\text{exp}}$ values for PbPb collisions at
2.76\TeV from Ref.~\cite{Giacalone:2016eyu} are also shown and found to be consistent with the current
measurements. Within the hydrodynamic model and allowing for a finite skewness of the event-by-event $v_2$ distribution,
the small splitting between the cumulant orders is expected to follow the relationship
($v_{2}\{6\}-v_{2}\{8\})/(v_{2}\{4\}-v_{2}\{6\}) = 0.091$~\cite{Giacalone:2016eyu}.
Experimentally, we find a value for this splitting ratio of $0.143 \pm 0.008\stat \pm 0.014\syst$ for 20--25\% central events,
with the ratio increasing to $0.185 \pm 0.005 \text{(stat)} \pm 0.012 \text{(syst)}$ as the centrality increases to 55--60\%.
The observed values might suggest higher order terms in a cumulant expansion of the $v_{2}$ distribution are required to account for the skewness.
This relationship was recently examined by the ALICE collaboration in Ref.~\cite{Acharya2018} using a q-cumulant analysis, with results comparable to the findings in this paper when considering systematic uncertainties and a different kinematic range for the ALICE measurement.

Both elliptic power and Bessel--Gaussian parametrizations used for fits such as shown in Fig.~\ref{fig:resultsUnf}  assume a linear response
between eccentricity and
flow, but only the elliptic power law allows for a finite skewness. For a Bessel-Gaussian distribution, the skewness is equal to zero.
This feature results in the elliptic power function being in better agreement with the observed fluctuation behavior than the Bessel--Gaussian parametrization, yielding $\chi^2/{\rm  dof}$ values on the order of unity. To avoid bin-to-bin correlations introduced by the unfolding procedure, goodness of fit values are obtained by refolding the fitted distributions with the response matrix and comparing to the measured distribution. The elliptic power $\chi^2/{\rm  dof}$ values vary between 0.8 and 1.5 from central to peripheral collisions, while the Bessel--Gaussian $\chi^2/{\rm  dof}$ values vary between 3 and 9.
 Point-by-point systematic uncertainties
on the unfolded distributions are correlated and are thus not considered in the fits.

\begin{figure*}[!htbp]
  \begin{center}
    \includegraphics[width=0.99\textwidth]{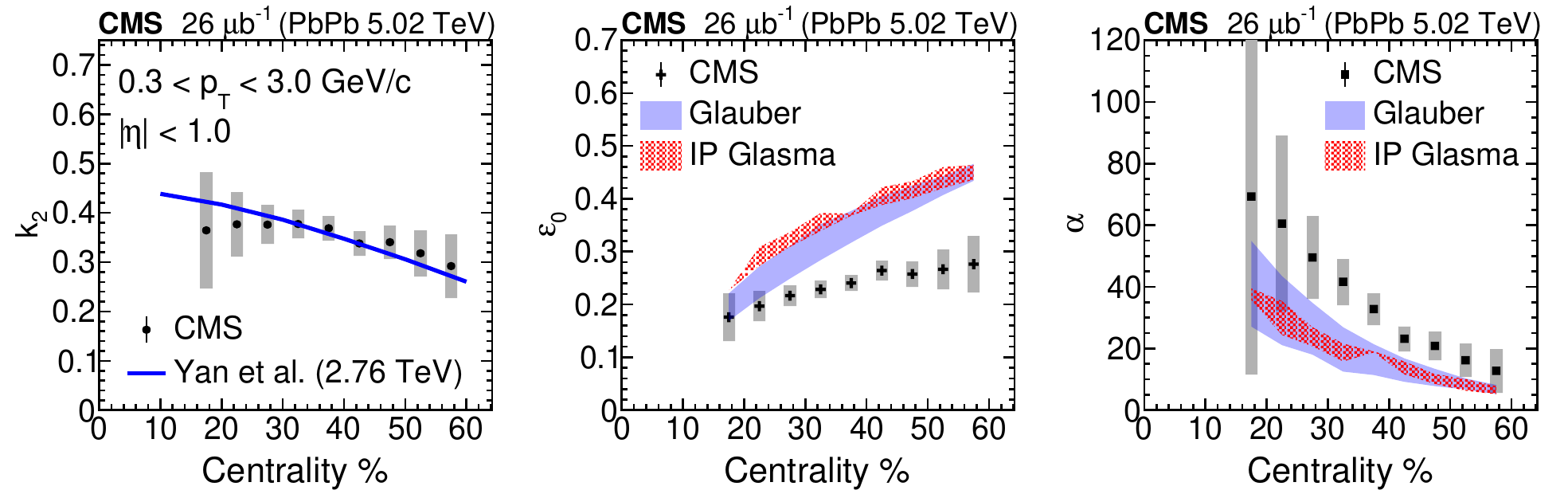}
    \caption{Centrality dependence of the parameters extracted from elliptic power function
             fits to the unfolded $p(v_2)$ distributions. Both statistical (error bars) and systematic (shaded boxes) uncertainties are shown.
             The solid line represents a theoretical calculation~\cite{Yan:2014nsa} using viscous hydrodynamics with Glauber
             initial conditions and an $\eta/s$ value of 0.19 to determine the
             response coefficient $k_2$. Glauber  (blue shaded band) and IP-Glasma (red shaded band) model calculations from
             Ref.~\cite{Yan:2014nsa} are shown for  the $\alpha$ and $\varepsilon_0$ parameters. The systematic uncertainties account for the highly correlated parameters of the elliptic power function fit and for the bin-to-bin correlations in the unfolded distributions introduced by the unfolding procedure.
             }
    \label{fig:EllParms}
  \end{center}
\end{figure*}

The fit parameters for the elliptic power function are shown in Fig.~\ref{fig:EllParms}
for the different centrality bins.  As also found in Ref.~\cite{Yan:2014nsa},  the fits do not converge for central collisions where the distributions become very close to a Bessel-Gaussian form.  Consequently, the parameters are shown for centralities $>$15\%.
The experimental $k_2$ values show only a weak centrality dependence.   Viscous hydrodynamic calculations indicate that deviations from thermal equilibrium should lead to a  reduced correspondence between the initial-state geometry and the flow signal in peripheral collisions~\cite{Voloshin:1999gs,Drescher:2007cd}.  This effect is suggested in Fig.~\ref{fig:EllParms} by the decrease in the $k_2$ value with increasing centrality, although the systematic uncertainties are too large for this to be a definitive observation.  The calculated decrease is greater than observed, although within the systematic uncertainties of the measurement. The eccentricity parameter of the power law fit, $\varepsilon_{0}$, is found to first increase, and then level off with increasing centrality. The leveling occurs for centralities $>40\%$, which is also where the $v_2$ values start to level off and then decrease. The $\alpha$ parameter, which reflects the number of sources in the power-law fit, is found to steadily decrease with increasing centrality, as expected.

Theoretical predictions at 2.76\TeV from Ref.~\cite{Yan:2014nsa} are compared to the
current analysis in Fig.~\ref{fig:EllParms}.
A viscous hydrodynamic calculation with Glauber initial conditions and an $\eta/s$ value of 0.19 is in agreement with the experimental $k_2$ values.
This coefficient is expected to have only a weak dependence on the initial state, with its centrality dependence largely determined by the viscosity of the medium~\cite{Yan:2014nsa}.
Predictions obtained using Glauber and
IP-Glasma~\cite{Schenke:2012fw,Schenke:2013aza} initial conditions, where the IP-Glasma model includes
gluon saturation effects, are shown for the  $\varepsilon_0$ and $\alpha$
parameters.  These latter two calculations  qualitatively capture the observed behavior for the  $\alpha$-parameter, but a significant difference is
found in comparing the theoretical $\varepsilon_0$ values with experiment. This difference might reflect a nonlinear response term, which will alter the
magnitude of the flow response coefficient and consequently the $\varepsilon_0$ and $\alpha$ parameters,
as suggested in Ref.~\cite{Yan:2014nsa}.

\section{Summary}
\label{sec:summary}
In summary, a non-Gaussian behavior is observed in the event-by-event fluctuations of the elliptic flow $v_2$
coefficients in PbPb collisions recorded by the CMS detector at $\rootsNN = 5.02$\TeV.
The probability distributions $p(v_2)$ for 5\%-centrality bins between 5\% and 60\% centrality
are found by unfolding statistical resolution effects from measured flow distributions. The
$v_2$ coefficients corresponding to different cumulant orders
are calculated from the moments of the unfolded $p(v_2)$
distributions. A rank ordering of  $v_2\{4\} > v_2\{6\} > v_2\{8\}$, with differences on the order of a few percent,
is observed for noncentral events with centralities greater than ${\approx}15\%$.
The standardized skewness of each $p(v_2)$ distribution is calculated using
the cumulant results.  In cases where there is a difference in the cumulant values,
the standardized skewness is found to be negative
with an increasing magnitude as collisions become less central.
Bessel--Gaussian and elliptic power functions are
fitted to the unfolded $p(v_2)$ distributions. The two distributions are similar for central collisions, though the
elliptic power function provides a better description for noncentral collisions.

Based on the elliptic power function fits, the centrality dependence of the
flow response coefficient, which relates the final state geometry to the initial
state energy density distribution, is found to be consistent with model calculations.
However, the observed eccentricities are smaller than predictions based on either the Glauber model
or the IP-Glasma model initial conditions with an assumed linear flow response. This difference might indicate the need for a nonlinear response term.
The current results illustrate that LHC experiments now have the precision to explore
the details of the initial-state fluctuations.

\begin{acknowledgments}
We congratulate our colleagues in the CERN accelerator departments for the excellent performance of the LHC and thank the technical and administrative staffs at CERN and at other CMS institutes for their contributions to the success of the CMS effort. In addition, we gratefully acknowledge the computing centers and personnel of the Worldwide LHC Computing Grid for delivering so effectively the computing infrastructure essential to our analyses. Finally, we acknowledge the enduring support for the construction and operation of the LHC and the CMS detector provided by the following funding agencies: BMWFW and FWF (Austria); FNRS and FWO (Belgium); CNPq, CAPES, FAPERJ, and FAPESP (Brazil); MES (Bulgaria); CERN; CAS, MoST, and NSFC (China); COLCIENCIAS (Colombia); MSES and CSF (Croatia); RPF (Cyprus); SENESCYT (Ecuador); MoER, ERC IUT, and ERDF (Estonia); Academy of Finland, MEC, and HIP (Finland); CEA and CNRS/IN2P3 (France); BMBF, DFG, and HGF (Germany); GSRT (Greece); OTKA and NIH (Hungary); DAE and DST (India); IPM (Iran); SFI (Ireland); INFN (Italy); MSIP and NRF (Republic of Korea); LAS (Lithuania); MOE and UM (Malaysia); BUAP, CINVESTAV, CONACYT, LNS, SEP, and UASLP-FAI (Mexico); MBIE (New Zealand); PAEC (Pakistan); MSHE and NSC (Poland); FCT (Portugal); JINR (Dubna); MON, RosAtom, RAS, RFBR and RAEP (Russia); MESTD (Serbia); SEIDI, CPAN, PCTI and FEDER (Spain); Swiss Funding Agencies (Switzerland); MST (Taipei); ThEPCenter, IPST, STAR, and NSTDA (Thailand); TUBITAK and TAEK (Turkey); NASU and SFFR (Ukraine); STFC (United Kingdom); DOE and NSF (USA).

\hyphenation{Rachada-pisek} Individuals have received support from the Marie-Curie program and the European Research Council and Horizon 2020 Grant, contract No. 675440 (European Union); the Leventis Foundation; the A. P. Sloan Foundation; the Alexander von Humboldt Foundation; the Belgian Federal Science Policy Office; the Fonds pour la Formation \`a la Recherche dans l'Industrie et dans l'Agriculture (FRIA-Belgium); the Agentschap voor Innovatie door Wetenschap en Technologie (IWT-Belgium); the Ministry of Education, Youth and Sports (MEYS) of the Czech Republic; the Council of Science and Industrial Research, India; the HOMING PLUS program of the Foundation for Polish Science, cofinanced from European Union, Regional Development Fund, the Mobility Plus program of the Ministry of Science and Higher Education, the National Science Center (Poland), contracts Harmonia 2014/14/M/ST2/00428, Opus 2014/13/B/ST2/02543, 2014/15/B/ST2/03998, and 2015/19/B/ST2/02861, Sonata-bis 2012/07/E/ST2/01406; the National Priorities Research Program by Qatar National Research Fund; the Programa Severo Ochoa del Principado de Asturias; the Thalis and Aristeia programs cofinanced by EU-ESF and the Greek NSRF; the Rachadapisek Sompot Fund for Postdoctoral Fellowship, Chulalongkorn University and the Chulalongkorn Academic into Its 2nd Century Project Advancement Project (Thailand); the Welch Foundation, contract C-1845; and the Weston Havens Foundation (USA).
\end{acknowledgments}

\bibliography{auto_generated}

\cleardoublepage \appendix\section{The CMS Collaboration \label{app:collab}}\begin{sloppypar}\hyphenpenalty=5000\widowpenalty=500\clubpenalty=5000\vskip\cmsinstskip
\textbf{Yerevan~Physics~Institute,~Yerevan,~Armenia}\\*[0pt]
A.M.~Sirunyan, A.~Tumasyan
\vskip\cmsinstskip
\textbf{Institut~f\"{u}r~Hochenergiephysik,~Wien,~Austria}\\*[0pt]
W.~Adam, F.~Ambrogi, E.~Asilar, T.~Bergauer, J.~Brandstetter, E.~Brondolin, M.~Dragicevic, J.~Er\"{o}, M.~Flechl, M.~Friedl, R.~Fr\"{u}hwirth\cmsAuthorMark{1}, V.M.~Ghete, J.~Grossmann, J.~Hrubec, M.~Jeitler\cmsAuthorMark{1}, A.~K\"{o}nig, N.~Krammer, I.~Kr\"{a}tschmer, D.~Liko, T.~Madlener, I.~Mikulec, E.~Pree, N.~Rad, H.~Rohringer, J.~Schieck\cmsAuthorMark{1}, R.~Sch\"{o}fbeck, M.~Spanring, D.~Spitzbart, W.~Waltenberger, J.~Wittmann, C.-E.~Wulz\cmsAuthorMark{1}, M.~Zarucki
% ...
\vskip\cmsinstskip
% ...
1:~Also at~Vienna~University~of~Technology,~Vienna,~Austria\\
% ...
\end{sloppypar}
\end{document}